\documentclass[a4paper,fleqn]{article} %Options in documentclass should be set to a4paper and fleqn.
\usepackage{modsim,bm}
\usepackage{times}
\usepackage{natbib} %The three packages modsim, times and natbib are required.
\usepackage{amsmath, amssymb, amsthm} %Also recommend the standard AMS LaTeX maths packages.

\MODSIMhead{M. C. Fung, G. W. Peters and P. V. Shevchenko, A State-Space Estimation of the Lee-Carter Mortality Model...} %This is the content of the headings in all pages (except the first page). The format should be author, title of paper. If the title is too long, then use ... at the end. If more than two authors, then please use the "et al." format with et al. in italic, for example A. Author {\it et al.}, Title of the paper.

\usepackage{rotating}
\usepackage{amsbsy,enumerate}
\usepackage{graphicx}
\usepackage{ccaption}
\usepackage{color}
\makeatletter
\renewcommand{\fnum@figure}[1]{\textbf{\figurename~\thefigure}. }
\renewcommand{\fnum@table}[1]{\textbf{\tablename~\thetable}. }
\makeatother

\begin{document}
% Defining Front matter.

\title{A State-Space Estimation of the Lee-Carter Mortality Model and Implications for Annuity Pricing}

\author{\underline{M. C. Fung} \address[A1]{\it{Risk Analytics Group, Digital Productivity Flagship, CSIRO, Australia}},
G. W. Peters \address[A2]{\it Department of Statistical Science,
University College London, United Kingdom\\
Associate Fellow of Oxford Mann Institute, Oxford University\\
Associate Fellow of Systemic Risk Center, London School of Economics.},
P. V. Shevchenko \address[A3]{{\it Risk Analytics Group, Digital Productivity Flagship, CSIRO, Australia} \\ {\bf{31 July 2015}}}} %underline the name of the presenting author. Use \addressmark[name-of-addressmark] after the name of an author who share the same affiliation (see modsim.tex for an example).

\email{Simon.Fung@csiro.au} %Email address of the presenting author only.

\date{Nov 2015}

%Keywords should be separated by commas and listed in Sentence case (first keyword with capital first letter and remaining keywords in lower case).

\begin{abstract}
A common feature of retirement income products is that their payouts depend on the lifetime of policyholders. A typical example is a life annuity policy which promises to provide benefits regularly as long as the retiree is alive. Consequently, insurers have to rely on ``best estimate" life tables, which consist of age-specific mortality rates, in order to price these kind of products properly. Recently there is a growing concern about the accuracy of the estimation of mortality rates since it has been historically observed that life expectancy is often underestimated in the past (so-called longevity risk), thus resulting in longer benefit payments than insurers have originally anticipated. To take into account the stochastic nature of the evolution of mortality rates, \cite{LeeCa92} proposed a stochastic mortality model which primarily aims to forecast age-specific mortality rates more accurately.

The original approach to estimating the Lee-Carter model is via a singular value decomposition, which falls into the least squares framework. Researchers then point out that the Lee-Carter model can
be treated as a state-space model. As a result several well-established state-space modeling techniques can be applied to not just perform estimation of the model, but to also perform forecasting as well as smoothing. Research in this area is still not yet fully explored in the actuarial literature, however. Existing relevant literature focuses mainly on mortality forecasting or pricing of longevity derivatives, while the full implications and methods of using the state-space representation of the Lee-Carter model in pricing retirement income products is yet to be
examined.

The main contribution of this article is twofold. First, we provide a rigorous and detailed derivation of the posterior distributions of the parameters and the latent process of the Lee-Carter model via Gibbs sampling. Our assumption for priors is slightly more general than the current literature in this area. Moreover, we suggest a new form of identification constraint not yet utilised in the actuarial literature that proves to be a more convenient approach for estimating the model under the state-space framework. Second, by exploiting the posterior distribution of the latent process and parameters, we examine the pricing range of annuities, taking into account the stochastic nature of the dynamics of the mortality rates. In this way we aim to capture the impact of longevity risk on the pricing of annuities.

The outcome of our study demonstrates that an annuity price can be more than $4\%$ under-valued when different assumptions are made on determining the survival curve constructed from the distribution of the forecasted mortality rates. Given that a typical annuity portfolio consists of a large number of policies with maturities which span decades, we conclude that the impact of longevity risk on the accurate pricing of annuities is a significant issue to be further researched. In addition, we find that mis-pricing is increasingly more pronounced for older ages as well as for annuity policies having a longer maturity.

\end{abstract} %Abstract should NOT extend beyond the first page.

\maketitle

\section{INTRODUCTION}
The pricing of retirement income products depends crucially on the accuracy of the predicted
death or survival probabilities. It is now widely documented that survival probability is
consistently underestimated especially in the last few decades (\cite{IMF12}). To capture the stochastic nature of mortality trends, \cite{LeeCa92} proposed a stochastic mortality model to forecast the trend of age-specific mortality rates.

There exists a body of literature on how to estimate the Lee-Carter model. The original approach in \cite{LeeCa92} is via singular value decomposition. To overcome the unrealistic feature of homogeneity in the additive error term, \cite{BrouhnsDeVe02} recast the model as a Poisson regression model assuming Poisson random variation for the number of deaths. Estimation of the model in the Poisson regression setting under the Bayesian framework is carried out in \cite{CzadoDeDe05}. Also there is a recently developed framework for modeling death counts with common risk factors via credit risk plus methodology and resultant estimation of the model via Monte Carlo Markov Chain in \cite{HirzScSh15}. In this paper we focus principally on the class of what has become known as the Lee-Carter models, in this regard another approach to estimating the Lee-Carter model is via state-space representation. \cite{Pedroza06} shows that the predictive intervals for forecasting are materially wider than using the singular value decomposition method. \cite{KogureKu10} adopt the state-space modeling approach and apply it for the pricing of longevity bonds and swaps.

In this paper we aim to explore further the Bayesian state-space modeling approach and examine its implication for annuity pricing. Specifically, we provide a rigorous and detailed derivation of the posterior distributions of the static parameters and the latent process of the Lee-Carter model via
Gibbs sampling. Our assumptions on the priors on the Lee-Carter model parameters are more general than \cite{Pedroza06} and \cite{KogureKu10}. Moreover, a new form of identification constraint not yet recognised in the actuarial literature is proposed which proves to be more convenient for estimating the model using an MCMC method under the state-space formulation. Using the predictive distributions of age-specific death rates, we examine the impact of longevity risk on the pricing of annuities and demonstrate that this long-term risk is indeed a significant factor when accurate pricing is required.

In Section~\ref{sec:LCModel} the state-space Lee-Carter model is presented together with some definitions and notation. Section~\ref{sec:BayesianLC} describes the Gibbs sampling approach to estimate the state-space Lee-Carter model. Posterior distributions of the static parameters and the latent process are derived in detail. Section~\ref{sec:pricing} examines the impact of longevity risk on annuity pricing. Section~\ref{sec:conclusion} concludes with some remarks.

\section{LEE-CARTER MODEL}\label{sec:LCModel}

\subsection{Definitions and Notation}\label{subsec:NotationsDefinitions}
%\subsubsection{Probability formulation}
In this section we briefly recall some important definitions from actuarial literature on mortality modelling that are required to set up the Lee-Carter model below and the pricing analysis in Section~\ref{sec:pricing}. We follow \cite{DicksonHaWa09} and \cite{PitaccoDeHaOl}. Let $T_x$ be a random variable representing the remaining lifetime
of a person aged $x$. The cumulative distribution function and survival function of $T_x$ are
written as ${}_{\tau}q_x = P(T_x \leq \tau)$ and ${}_{\tau}p_x = P(T_x > \tau)$ respectively. For a person aged $x$, the force of mortality at age $x+\tau$  is defined as
\begin{equation}\label{eqn:forcemortality}
    \mu_{x+\tau} := \lim_{h\rightarrow 0} \frac{1}{h}P(T_x<\tau+h|T_x>\tau)
    %&= \lim_{h\rightarrow 0} \frac{1}{h}\frac{P(t < T_x <
    %t+h)}{P(T_x>t)} \\
    = \frac{1}{{}_{\tau}p_x}\lim_{h\rightarrow
    0}\frac{1}{h}({}_{\tau+h}q_x - {}_{\tau}q_x)
    = \frac{1}{{}_{\tau}p_x}\frac{d}{d\tau}{}_{\tau}q_x %\label{eq:mu}
    = -\frac{d}{d\tau}\ln{{}_{\tau}p_x}
\end{equation}
and hence
%\begin{equation}\label{eq:survProb}
$    {}_{\tau}p_x = \exp{\left(-\int^\tau_0 \mu_{x+s}\,ds\right)}$.
%\end{equation}
Note that the survival probability function has the following
important property: ${}_{\tau+u}p_x = {}_{\tau}p_{x+u} \times {}_{u}p_x $.
Let $f_x(t)$ be the density function of $T_x$, then from \eqref{eqn:forcemortality} we see that
${}_{\tau}q_x = \int^\tau_0 f_x(s)\,ds =\int^\tau_0 {}_{s}p_x\,\mu_{x+s}\,ds$.
The central death rate for a $x$-year-old, where
$x \in \mathbb{N}$, is defined as
\begin{equation}\label{eqn:centralmortality}
m_x := \frac{{}_1q_x}{\int^1_0 {}_{s}p_x\,ds} = \frac{\int^1_0
{}_{s}p_x\,\mu_{x+s}\,ds}{\int^1_0 {}_{s}p_x\,ds}
\end{equation}
which is a weighted-average of the force of
mortality. Under the so-called piecewise constant force of mortality assumption, that is $\mu_{x+s} = \mu_x$ where
$0 \leq s<1$ and $x \in \mathbb{N}$, we have, from \eqref{eqn:centralmortality}, $m_x = \mu_x$ and hence ${}_1p_x = e^{-m_x}$. Moreover, the maximum likelihood estimate of the force of mortality $\hat{\mu}_x$ (and hence $\hat{m}_x$) is given by $\hat{\mu}_x=D_x/E_x=\hat{m}_x$ where $D_x$ is the number of deaths recorded at age $x$ last birthday and the exposure-to-risk $E_x$ is the total time lived by people aged $x$ last birthday, during the observation year. Note that $E_x$ is often approximated by an estimate of the population aged $x$ last birthday in the middle of the observation year.

\subsection{The Lee-Carter State Space Model}\label{subsec:originalLC}

Based on the definitions described above, we now discuss the work of \citet{LeeCa92} who proposed a stochastic mortality model specifically for forecasting age-specific central death rates $m_{xt}$, where $x=x_1,\dots,x_p$ and $t=1,\dots,n$ represent age and year (time) respectively. The model assumes that the log central death rate, $y_{xt}=\ln{m_{xt}}$, is governed by the following equation
\begin{equation}
    \boldsymbol{y}_t = \boldsymbol{\alpha}+\boldsymbol{\beta}\kappa_t+ \boldsymbol{\varepsilon}_t,
    \quad \boldsymbol{\varepsilon}_t \sim \text{N}(0,\sigma^2_\varepsilon \boldsymbol{1}_p) \label{eqn:LCy} \\
\end{equation}
where $\boldsymbol{y}_t=(y_{x_1t},\dots,y_{x_pt})'$,
$\boldsymbol{\alpha}=(\alpha_{x_1},\dots,\alpha_{x_p})'$,
$\boldsymbol{\beta}=(\beta_{x_1},\dots,\beta_{x_p})'$,
$\boldsymbol{\varepsilon}_t=(\varepsilon_{x_1t},\dots,\varepsilon_{x_pt})'$, $\boldsymbol{1}_p$ is the $p$ by $p$ identity matrix and $\text{N}(.,.)$ denotes the Gaussian distribution. \citet{LeeCa92} estimate the model \eqref{eqn:LCy} via singular value decomposition and subsequently assume that the unobserved latent time trend denoted by $\kappa_t$ satisfies the following linear dynamics
\begin{equation}
    \kappa_t = \kappa_{t-1}+\theta+\omega_{t}, \quad \omega_t \sim
    \text{N}(0,\sigma^2_\omega) \label{eqn:LCk}
\end{equation}
where $\boldsymbol{\varepsilon}_t$ and $\omega_t$ are independent. The parameters $\theta, \sigma^2_\omega$ are then estimated using standard econometric techniques. In this form the Lee-Carter model is, however, not identifiable since the model \eqref{eqn:LCy} is invariant up to some linear transformations of the parameters:
\begin{equation}\label{eqn:LCtransform}
\boldsymbol{y}_t = \boldsymbol{\alpha}+\boldsymbol{\beta}\kappa_t+
\boldsymbol{\varepsilon}_t  = \boldsymbol{\alpha}+
\boldsymbol{\beta}c +
\frac{\boldsymbol{\beta}}{d}\left((\kappa_t-c)d\right)+\boldsymbol{\varepsilon}_t
=
\tilde{\boldsymbol{\alpha}}+\tilde{\boldsymbol{\beta}}\tilde{\kappa}_t+
\boldsymbol{\varepsilon}_t
\end{equation}
where $\tilde{\boldsymbol{\alpha}}=\boldsymbol{\alpha}+
\boldsymbol{\beta}c$,
$\tilde{\boldsymbol{\beta}}=\frac{\boldsymbol{\beta}}{d}$ and
$\tilde{\kappa}_t=(\kappa_t-c)d$. To overcome this identification issue, \citet{LeeCa92} introduced the following constraints
\begin{equation}\label{eqn:constraints}
    \sum^{x_p}_{x=x_1}\beta_x=1, \quad \sum^n_{t=1}\kappa_t=0
\end{equation}
to ensure that the model becomes identifiable since, by setting $d = \sum^{x_p}_{x=x_1} \beta_x$ and $c = \sum^n_{t=1} \kappa_t$, we have
$\sum^{x_p}_{x=x_1}\tilde{\beta}_x=1$ and $\sum^n_{t=1}\tilde{\kappa}_t=0$.

\cite{Pedroza06} suggests that we can in
fact combine the processes $\boldsymbol{y}_t$ and $\kappa_t$ into
one dynamical system
\begin{equation}\label{eqn:LCstatespace}
    \boldsymbol{y}_t = \boldsymbol{\alpha}+\boldsymbol{\beta}\kappa_t+ \boldsymbol{\varepsilon}_t,
    \quad
    \kappa_t = \kappa_{t-1}+\theta+\omega_{t},\text{ where}\quad  \boldsymbol{\varepsilon}_t \sim \text{N}(0,\sigma^2_\varepsilon \boldsymbol{1}_p),
    \quad \omega_t \sim
    \text{N}(0,\sigma^2_\omega)
\end{equation}
%\begin{align}
%    \boldsymbol{y}_t &= \boldsymbol{\alpha}+\boldsymbol{\beta}\kappa_t+ \boldsymbol{\varepsilon}_t,
%    \quad \boldsymbol{\varepsilon}_t \sim \text{N}(0,\sigma^2_\varepsilon \boldsymbol{1}_p) \label{eqn:LCStateSpace1} \\
%    \kappa_t &= \kappa_{t-1}+\theta+\omega_{t}, \quad \omega_t \sim
%    \text{N}(0,\sigma^2_\omega), \label{eqn:LCStateSpace2}
%\end{align}
resulting in a state-space representation of the Lee-Carter model and estimate $\kappa_t$ and model parameters jointly.

\textbf{Note on Estimation via Lee-Carter Framework:} although Lee-Carter model is expressed in a state-space formulation, given this form of the identification constraints it is not readily amenable to standard state-space estimation procedures since the constraint is expressed on the path-space of the latent process. Consequently, this led \cite{LeeCa92} to develop an alternative estimation procedure where the first step in the estimation does not depend on the dynamics of $\kappa_t$and utilises a singular value decomposition (SVD) approach. Then the evolution of $\kappa_t$ is specified after the SVD procedure is performed, meaning that although the state space structure is specified, this form of the representation is not exploited in the estimation of the model trend $\kappa_t$ or the model parameters. We will demonstrate in this paper how to change the identification constraints so that standard filtering based state-space model estimation procedures can be utilised.

\subsection{Lee-Carter model in ARIMA Time Series Form}
We also note that, at least when one doesn't consider the identification constraints, the Lee-Carter model is a simple linear dynamic model. Hence, we also highlight that this model can be rewritten in the form of an ARIMA structure via a Local Level formulation where we denote $\boldsymbol{\eta}_t := \boldsymbol{\alpha} + \boldsymbol{\beta}\kappa_t$ and $h_t := \theta + w_t$. One can then rewrite the state-space form where each age $x$ is an ARIMA(0,1,1) structure as $\boldsymbol{Z}_{xt}:= \nabla \boldsymbol{Y}_{xt} = \nabla \boldsymbol{\eta}_{xt} + \nabla \boldsymbol{\epsilon}_{xt}$ with a simple closed form expression for the auto-correlation function given by
\begin{equation}
\rho_{Z_{x}}(k) = \frac{\gamma(k)}{\gamma(0)} =
\begin{cases}
-\frac{\sigma_{\epsilon}^2}{\sigma_{w}^2 + 2\sigma_{\epsilon}^2}, & k = 1,\\
0, & k \geq 1.
\end{cases}
\end{equation}
Suggesting that one can also perform estimation on the unconstrained form of the model via estimation based on the autocorrelation, though these would need to be modified subject to identification constraints. This would again complicate the estimation, suggesting the need to try to find alternative identification constraints that are more applicable to these standard estimation approaches.

\section{BAYESIAN INFERENCE FOR LEE-CARTER MODEL IN STATE-SPACE FORM} \label{sec:BayesianLC}

\cite{Pedroza06} and \cite{KogureKu10} both consider Bayesian formulations of the Lee-Carter model which allows the joint estimation of $\kappa_t$ and model parameters. However, under their formulation they again work with the identification constraints of \eqref{eqn:constraints} which are not obvious to use when designing efficient Monte Carlo procedures such as an Markov Chain Monte Carlo (MCMC) procedure. Such identification constraints will lead to difficulties in designing the proposal of the MCMC and difficulties in achieving suitable acceptance rates for the resultant Markov chain, resulting in high variance in estimates of mortality rates.

Additionally, although these authors work in the Bayesian setting, their derivations of the posterior distributions are not fully described. In the following we derive the posterior distributions of the parameters and the state process of the Lee-Carter model under our extended Bayesian framework.

\subsection{Lee-Carter model: New Identification Constraints and Bayesian Formulations}
In this article, we suggest an alternative new formulation of the identification constraints required which we believe is simpler and more readily applicable to most Monte Carlo based procedures such as MCMC and filtering methods such as Kalman Filter and Sequential Monte Carlo. This has the key advantage that for a given computational effort we can design efficient MCMC samplers with lower variance and therefore result in more reliable estimates of mortality rate. Our formulation of the identification constraints are given by simply setting $\alpha_{x_1} = \text{constant}$, and $\beta_{x_1} = \text{constant}$. Such a choice is a valid identification constraint since if one of the elements of each $\boldsymbol{\alpha}$ and $\boldsymbol{\beta}$ are known, then a non-trivial linear transformation in \eqref{eqn:LCtransform} is not allowed anymore; that is, we must have $c=0$ and $d=1$.

Under the Bayesian approach, we aim to obtain the posterior density $\pi(\kappa_{0:n},\boldsymbol{\Psi}|\boldsymbol{y}_{1:n})$ of the states \footnote{Here $a_{1:t}$ means $a_1,\dots,a_t$.}
$\kappa_{0:n}$ as well as the parameters, $\boldsymbol{\Psi} :=(\alpha_{x_2:x_p},\beta_{x_2:x_p},\theta,\sigma^2_\varepsilon,\sigma^2_\omega)$, given the observations $\boldsymbol{y}_{1:n}$. Note that $\alpha_{x_1}$ and $\beta_{x_1}$ are assumed to be known constraints. Under such a Bayesian formulation, it is standard to utilise a MCMC procedure to sample from $\pi(\kappa_{0:n},\boldsymbol{\Psi}|\boldsymbol{y}_{1:n})$, see discussions on such procedures in risk and insurance settings in \cite{CruzPeSh15}.

In this paper we explain an efficient and suitable sampling framework for actuarial applications which utilises the state-space Lee-Carter structure, in particular the fact that it is a linear Gaussian model, as well as the new constraint formulation we introduce. Under this model we develop an efficient approach involving a combined Gibbs sampling conjugate model sampler for the marginal target distributions of the static model parameters along with a forward backward Kalman filter sampler for the latent process $\kappa_{1:t}$.

A sample of the targeted density is obtained via Gibbs sampling in two steps: (1) Initialise $\boldsymbol{\Psi}=\boldsymbol{\Psi}^{(0)}$; (2) For $i=1,\dots,N$, first draw $\kappa^{(i)}_{0:n}$ from
    $\pi(\kappa_{0:n}|\boldsymbol{\Psi}^{(i-1)},\boldsymbol{y}_{1:n})$, then draw $\boldsymbol{\Psi}^{(i)}$ from
    $\pi(\boldsymbol{\Psi}|\kappa^{(i)}_{0:n},\boldsymbol{y}_{1:n})$.

\subsection{Sampling from the full conditional density $\pi(\kappa_{0:n}|\boldsymbol{\Psi},\boldsymbol{y}_{1:n})$}
Samples from the full conditional density $\pi(\kappa_{0:n}|\boldsymbol{\Psi},\boldsymbol{y}_{1:n})$ can be obtained via the so-called forward-filtering-backward sampling (FFBS) procedure (\cite{CarterKo94}). We can write
\begin{equation}\label{eqn:ffbs}
    \pi(\kappa_{0:n}|\boldsymbol{\Psi},\boldsymbol{y}_{1:n})=\prod^n_{t=0}\pi(\kappa_t|\kappa_{t+1:n},\boldsymbol{\Psi},\boldsymbol{y}_{1:n})=
    \prod^n_{t=0}\pi(\kappa_t|\kappa_{t+1},\boldsymbol{\Psi},\boldsymbol{y}_{1:t})
\end{equation}
where the last term in the product, $\pi(\kappa_n|\boldsymbol{\Psi},\boldsymbol{y}_{1:n})$, is distributed as $\text{N}(m_n,C_n)$ in Kalman filtering. We use the following notation
\begin{align}
    \kappa_{t-1}|\boldsymbol{y}_{1:t-1} &\sim \text{N}(m_{t-1},C_{t-1})\\
    \kappa_t|\boldsymbol{y}_{1:t-1} &\sim \text{N}(a_t, R_t), \text{ where}\quad  a_t=m_{t-1}+\theta, R_t=C_{t-1}+\sigma^2_\omega \label{eqn:kf1} \\
    \boldsymbol{y}_t|\boldsymbol{y}_{1:t-1} &\sim \text{N}(\boldsymbol{f}_t,\boldsymbol{Q}_t), \text{ where}\quad
        \boldsymbol{f}_t=\boldsymbol{\alpha}+\boldsymbol{\beta} a_t, \boldsymbol{Q}_t=\boldsymbol{\beta}\boldsymbol{\beta}' R_t+\sigma^2_\varepsilon \boldsymbol{1}_p    \\
    \kappa_t|\boldsymbol{y}_{1:t} &\sim \text{N}(m_t,C_t), \text{ where}\quad
        m_t=a_t+R_t\boldsymbol{\beta}'\boldsymbol{Q}_t^{-1}(\boldsymbol{y}_t-\boldsymbol{f}_t),
        C_t=R_t-R_t\boldsymbol{\beta}'\boldsymbol{Q}_t^{-1}\boldsymbol{\beta}R_t \label{eqn:kf4}
\end{align}
to denote the distributions involved in Kalman filtering. Once we draw a sample $\kappa_n$ from $\text{N}(m_n,C_n)$, then Eq.~\eqref{eqn:ffbs} suggests that we can draw recursively and
backwardly $\kappa_t$ from $\pi(\kappa_t|\kappa_{t+1},\boldsymbol{\Psi},\boldsymbol{y}_{1:t})$
where $t=n-1,n-2,\dots,1,0$. It can be shown that (\cite{PetrisPeCa})
\begin{equation}
    \pi(\kappa_t|\kappa_{t+1},\boldsymbol{\Psi},\boldsymbol{y}_{1:t}) \sim \text{N}(h_t,H_t),\text{ where}\quad
    h_t = m_t+C_tR^{-1}_{t+1}(\kappa_{t+1}-a_{t+1}), H_t =
    C_t-C_tR^{-1}_{t+1}C_t.
\end{equation}
In summary, the FFBS algorithm consists of three steps: (1) Run Kalman filter to obtain $m_n$ and $C_n$; (2) Draw $\kappa_n$ from $\text{N}(m_n,C_n)$ and (3) For $t=n-1,\dots,0$, draw $\kappa_t$ from $\text{N}(h_t,H_t)$.

\subsection{Sampling from the full conditional desnity $\pi(\boldsymbol{\Psi}|\kappa_{0:n},\boldsymbol{y}_{1:n})$}
Sampling from the full conditional density $\pi(\boldsymbol{\Psi}|\kappa_{0:n},\boldsymbol{y}_{1:n})$ can be achieved by applying Gibbs sampling. The prior
for $(\alpha_x, \beta_x, \theta, \sigma^2_\varepsilon, \sigma^2_\omega)$ are given by
$\alpha_x \sim \text{N}(\tilde{\mu}_\alpha,\tilde{\sigma}^2_\alpha),
        \text{ } \beta_x \sim
        \text{N}(\tilde{\mu}_\beta,\tilde{\sigma}^2_\beta),
        \text{ } \sigma^2_{\varepsilon} \sim
        \text{IG}(\tilde{a}_\varepsilon, \tilde{b}_\varepsilon), \text{ }
    \theta \sim
        \text{N}(\tilde{\mu}_\theta,
        \tilde{\sigma}^2_\theta), \text{ }
        \sigma^2_\omega \sim
        \text{IG}(\tilde{a}_\omega, \tilde{b}_\omega)$
where $x \in \{x_2,\dots,x_p\}$ and $\text{IG}(.,.)$ denote the inverse-gamma
distribution. It is assumed that the priors for all parameters are independent. In this case the posterior densities of parameters are of the same type as the prior densities, a so-called conjugate prior. In the following we derive the posterior distribution for each parameter (for ease of notation it is assumed that $\boldsymbol{y}=\boldsymbol{y}_{1:n}$, $\boldsymbol{\kappa}=\kappa_{0:n}$,
family $\boldsymbol{\Psi}_{-\lambda}$ means ``$\boldsymbol{\Psi}$ without the parameter $\lambda$"):
\begin{itemize}
  \item For $\alpha_x$ where $x \in \{x_2,\dots,x_p\}$, we have
        \begin{align} \notag
            \pi(\alpha_x|\boldsymbol{y},\boldsymbol{\kappa}, \boldsymbol{\Psi}_{-\alpha_x})
            &\propto
            \pi(\boldsymbol{y}|\boldsymbol{\kappa},\boldsymbol{\Psi})\pi(\boldsymbol{\kappa}|\boldsymbol{\Psi})
            \pi(\alpha_{x}|\boldsymbol{\Psi}_{-\alpha_x})
            \propto
            \prod^n_{t=1}\pi(y_{xt}|\kappa_t,\alpha_x,\beta_x,\sigma^2_\varepsilon)\pi(\alpha_x) \notag \\
            %&\propto
            %\exp\left\{-\frac{1}{2\tilde{\sigma}^2_\alpha}(\alpha_x-\tilde{\mu}_\alpha)^2-\frac{1}{2\sigma^2_\varepsilon}\sum^n_{t=1}(y_{xt}-\alpha_x-\beta_x\kappa_t)^2\right\} \notag \\
            &\propto
            \exp\left\{-\frac{1}{2}\left(\frac{(\tilde{\sigma}^2_\alpha n+\sigma^2_\varepsilon)\alpha^2_x-2(\tilde{\mu}_\alpha\sigma^2_\varepsilon+ \tilde{\sigma}^2_\alpha\sum_t(y_{xt}-\beta_x\kappa_t))\alpha_x}
            {\tilde{\sigma}^2_\alpha\sigma^2_\varepsilon}\right)\right\}. \notag
        \end{align}
    Hence the posterior conditional distribution of $\alpha_x$ is given by $\text{N}\left(\frac{\tilde{\mu}_\alpha\sigma^2_\varepsilon+\tilde{\sigma}^2_\alpha\sum_t
            (y_{xt}-\beta_x\kappa_t)}{\tilde{\sigma}^2_\alpha n+\sigma^2_\varepsilon},
            \frac{\tilde{\sigma}^2_\alpha\sigma^2_\varepsilon}{\tilde{\sigma}^2_\alpha n+\sigma^2_\varepsilon}\right)$.
        %\begin{equation}\label{posterior_alpha}
        %    \pi(\alpha_x|\boldsymbol{y},\kappa, \boldsymbol{\Psi}_{-\alpha_x}) \sim %\text{N}\left(\frac{\tilde{\mu}_\alpha\sigma^2_\varepsilon+\tilde{\sigma}^2_\alpha\sum_t
        %    (y_{xt}-\beta_x\kappa_t)}{\tilde{\sigma}^2_\alpha n+\sigma^2_\varepsilon},
        %    \frac{\tilde{\sigma}^2_\alpha\sigma^2_\varepsilon}{\tilde{\sigma}^2_\alpha n+\sigma^2_\varepsilon}\right)
        %\end{equation}
  \item For $\beta_x$ where $x \in \{x_2,\dots,x_p\}$, we have
        \begin{align}
            \pi(\beta_x|\boldsymbol{y},\boldsymbol{\kappa}, \boldsymbol{\Psi}_{-\beta_x})
            &\propto
            \pi(\boldsymbol{y}|\boldsymbol{\Psi})\pi(\boldsymbol{\kappa}|\boldsymbol{\Psi})
            \pi(\beta_{x}|\boldsymbol{\Psi}_{-\beta_x})
            \propto
            \prod^n_{t=1}\pi(y_{xt}|\kappa_t,\alpha_x,\beta_x,\sigma^2_\varepsilon)\pi(\beta_x) \notag\\
            %&\propto
            %\exp\left\{-\frac{1}{2\tilde{\sigma}^2_\beta}(\beta_x-\tilde{\mu}_\beta)^2-\frac{1}{2\sigma^2_\varepsilon}\sum^n_{t=1}(y_{xt}-\alpha_x-\beta_x\kappa_t)^2\right\}\notag \\
            &\propto
            \exp\left\{-\frac{1}{2}\left(\frac{(\tilde{\sigma}^2_\beta\sum_t\kappa^2_t+\sigma^2_\varepsilon)\beta^2_x-2\left(\tilde{\mu}_\beta\sigma^2_\varepsilon+
            \tilde{\sigma}^2_\beta\sum_t(y_{xt}-\alpha_x)\kappa_t\right)\beta_x}
            {\tilde{\sigma}^2_\beta\sigma^2_\varepsilon}\right)\right\}.\notag
        \end{align}
    Hence the posterior conditional distribution of $\beta_x$ is given by $\text{N} \left(\frac{\tilde{\sigma}^2_\beta\sum_t(y_{xt}-\alpha_x)\kappa_t+\tilde{\mu}_\beta\sigma^2_\varepsilon}
            {\tilde{\sigma}^2_\beta\sum_t\kappa^2_t+\sigma^2_\varepsilon},
            \frac{\tilde{\sigma}^2_\beta\sigma^2_\varepsilon}{\tilde{\sigma}^2_\beta\sum_t\kappa^2_t+\sigma^2_\varepsilon}\right)$.
        %\begin{equation}
        %    \pi(\beta_x|\boldsymbol{y},\kappa, \boldsymbol{\Psi}_{-\beta_x}) \sim \text{N} %\left(\frac{\tilde{\sigma}^2_\beta\sum_t(y_{xt}-\alpha_x)\kappa_t+\tilde{\mu}_\beta\sigma^2_\varepsilon}
        %    {\tilde{\sigma}^2_\beta\sum_t\kappa^2_t+\sigma^2_\varepsilon},
        %    \frac{\tilde{\sigma}^2_\beta\sigma^2_\varepsilon}{\tilde{\sigma}^2_\beta\sum_t\kappa^2_t+\sigma^2_\varepsilon}\right)
        %\end{equation}
  \item For $\theta$, we have
        \begin{align}
            \pi(\theta|\boldsymbol{y},\boldsymbol{\kappa},\boldsymbol{\Psi}_{-\theta})
            &\propto
            \pi(\boldsymbol{y}|\boldsymbol{\kappa},\boldsymbol{\Psi})\pi(\boldsymbol{\kappa}|\boldsymbol{\Psi})
            \pi(\theta|\boldsymbol{\Psi}_{-\theta})
            \propto
            \prod^n_{t=1}\pi(\kappa_t|\kappa_{t-1},\theta,\sigma^2_\omega)\pi(\theta) \notag \\
            %&\propto
            %\exp\left\{-\frac{1}{2\tilde{\sigma}^2_\theta}(\theta-\tilde{\mu}_\theta)^2-\frac{1}{2\sigma^2_\omega}\sum^n_{t=1}(\kappa_t-(\kappa_{t-1}+\theta))^2\right\} \notag \\
            &\propto
            \exp\left\{-\frac{1}{2}\left(\frac{(\tilde{\sigma}^2_\theta n+\sigma^2_\omega)\theta^2-2\left(\tilde{\mu}_\theta\sigma^2_\omega+
            \tilde{\sigma}^2_\theta\sum_t(\kappa_t-\kappa_{t-1})\right)\theta}
            {\tilde{\sigma}^2_\theta\sigma^2_\omega}\right)\right\}.  \notag
        \end{align}
    Hence the posterior conditional distribution of $\theta$ is given by $\text{N} \left(\frac{\tilde{\sigma}^2_\theta\sum^n_{t=1}(\kappa_t-\kappa_{t-1})+\tilde{\mu}_\theta\sigma^2_\omega}
            {\tilde{\sigma}^2_\theta n+\sigma^2_\omega},
            \frac{\tilde{\sigma}^2_\theta \sigma^2_\omega}{\tilde{\sigma}^2_\theta n+\sigma^2_\omega}\right)$.
        %\begin{equation} \label{posterior_theta}
        %    \pi(\theta|\boldsymbol{y},\kappa,\boldsymbol{\Psi}_{-\theta}) \sim \text{N} %\left(\frac{\tilde{\sigma}^2_\theta\sum^n_{t=1}(\kappa_t-\kappa_{t-1})+\tilde{\mu}_\theta\sigma^2_\omega}
        %    {\tilde{\sigma}^2_\theta n+\sigma^2_\omega},
        %    \frac{\tilde{\sigma}^2_\theta \sigma^2_\omega}{\tilde{\sigma}^2_\theta n+\sigma^2_\omega}\right)
        %\end{equation}
  \item For $\sigma^2_\varepsilon$, we have
    \begin{align}
            \pi(\sigma^2_\varepsilon|\boldsymbol{y},\boldsymbol{\kappa},\boldsymbol{\Psi}_{-\sigma^2_\varepsilon})
            &\propto
            \pi(\boldsymbol{y}|\boldsymbol{\kappa},\boldsymbol{\Psi})\pi(\boldsymbol{\kappa}|\boldsymbol{\Psi})
            \pi(\sigma^2_\varepsilon|\boldsymbol{\Psi}_{-\sigma^2_\varepsilon})
            \propto
            \prod^n_{t=1}\prod^{x_p}_{x=x_1}\pi(y_{xt}|\kappa_t,\alpha_x,\beta_x,\sigma^2_\varepsilon)\pi(\sigma^2_\varepsilon)\notag \\
            %&\propto
            %\frac{1}{(\sigma^2_\varepsilon)^{np/2}}\exp\left\{-\frac{1}{2\sigma^2_\varepsilon}
            %\sum_t\sum_x\left(y_{xt}-(\alpha_x+\beta_x\kappa_t)\right)^2\right\}\frac{1}{(\sigma^2_\varepsilon)^{\tilde{a}_\varepsilon+1}}
            %\exp\left\{-\frac{\tilde{b}_\varepsilon}{\sigma^2_\varepsilon}\right\} \notag \\
            &\propto
            \frac{1}{(\sigma^2_\varepsilon)^{np/2+\tilde{a}_\varepsilon+1}}\exp\left\{-\frac{1}{\sigma^2_\varepsilon}\left(\tilde{b}_\varepsilon+\frac{1}{2}
            \sum_t\sum_x\left(y_{xt}-(\alpha_x+\beta_x\kappa_t)\right)^2\right)\right\}.\notag
        \end{align}
    The posterior conditional distribution of $\sigma^2_\varepsilon$ is thus $\text{IG} \left(\tilde{a}_\varepsilon+\frac{np}{2},\,
            \tilde{b}_\varepsilon+\frac{1}{2}
            \sum^n_{t=1}\sum^{x_p}_{x=x_1}\left(y_{xt}-(\alpha_x+\beta_x\kappa_t)\right)^2\right)$.
        %\begin{equation}
        %    \pi(\sigma^2_\varepsilon|\boldsymbol{y},\kappa, \boldsymbol{\Psi}_{-\sigma^2_\varepsilon}) \sim \text{IG} %\left(\tilde{a}_\varepsilon+\frac{np}{2},\,
        %    \tilde{b}_\varepsilon+\frac{1}{2}
        %    \sum^n_{t=1}\sum^{x_p}_{x=x_1}\left(y_{xt}-(\alpha_x+\beta_x\kappa_t)\right)^2\right)
        %\end{equation}
  \item For $\sigma^2_\omega$, we have
    \begin{align}
            \pi(\sigma^2_\omega|\boldsymbol{y},\boldsymbol{\kappa}, \boldsymbol{\Psi})
            &\propto
            \pi(\boldsymbol{y}|\boldsymbol{\kappa},\boldsymbol{\Psi})\pi(\boldsymbol{\kappa}|\boldsymbol{\Psi})
            \pi(\sigma^2_\omega|\boldsymbol{\Psi}_{-\sigma^2_\omega})
            \propto
            \prod^n_{t=1}\pi(\kappa_t|\kappa_{t-1},\theta,\sigma^2_\omega)\pi(\sigma^2_\omega)\notag \\
            %&\propto
            %\frac{1}{(\sigma^2_\omega)^{n/2}}\exp\left\{-\frac{1}{2\sigma^2_\omega}
            %\sum_t\left(\kappa_t-(\kappa_{t-1}+\theta)\right)^2\right\}\frac{1}{(\sigma^2_\omega)^{\tilde{a}_\omega}+1}
            %\exp\left\{-\frac{\tilde{b}_\omega}{\sigma^2_\omega}\right\}\notag \\
            &\propto
            \frac{1}{(\sigma^2_\omega)^{n/2+\tilde{a}_\omega+1}}\exp\left\{-\frac{1}{\sigma^2_\omega}\left(\tilde{b}_\omega+\frac{1}{2}
            \sum_t\left(\kappa_t-(\kappa_{t-1}+\theta)\right)^2\right)\right\}.\notag
        \end{align}
    The posterior conditional distribution of $\sigma^2_\omega$ is
    thus $\text{IG} \left(\tilde{a}_\omega+\frac{n}{2},\,
            \tilde{b}_\omega+\frac{1}{2}
            \sum^n_{t=1}\left(\kappa_t-(\kappa_{t-1}+\theta)\right)^2\right).$
        %\begin{equation}
        %    \pi(\sigma^2_\omega|\boldsymbol{y},\kappa,\boldsymbol{\Psi}) \sim \text{IG} \left(\tilde{a}_\omega+\frac{n}{2},\,
        %    \tilde{b}_\omega+\frac{1}{2}
        %    \sum^n_{t=1}\left(\kappa_t-(\kappa_{t-1}+\theta)\right)^2\right).
        %\end{equation}
\end{itemize}

\subsection{Forecasting}
%If parameters are known, the predictions $\kappa_{n+k}$ and $\boldsymbol{y}_{n+k}$, given $\boldsymbol{y}_{n}$, are %Gaussian distributed and hence can be obtained by calculating the corresponding mean and variance using iterated %expectation, see \cite{PetrisPeCa}. In the case when parameters are random variables which are sampled using MCMC, we
The predictive distributions of $\boldsymbol{y}_{n+k}$, given $\boldsymbol{y}_{n}$, are obtained using the MCMC samples as follows. Let $L$ be the number of samples remained after burn-in. Then for $k \geq 1$, and for $\ell=1\dots,L$, we sample recursively
\begin{equation}
    \kappa_{n+k}^{(\ell)} \sim \text{N}\left(\kappa^{(\ell)}_{n+k-1}+\theta^{(\ell)},\left(\sigma^2_\omega\right)^{(\ell)}\right),\quad
    \boldsymbol{y}^{(\ell)}_{n+k} \sim \text{N}\left(\boldsymbol{\alpha}^{(\ell)}+\boldsymbol{\beta}^{(\ell)}\kappa^{(\ell)}_{n+k},
        \left(\sigma^2_\varepsilon\right)^{(\ell)}\boldsymbol{1}_p\right)
\end{equation}
where the samples $\kappa^{(\ell)}_n$ are obtained from the FFBS procedure. This produces an estimate of $\pi(\boldsymbol{y}_{n+k}|\boldsymbol{y}_{1:n})=\int \pi(\boldsymbol{y}_{n+k}|\kappa_{n+k},\boldsymbol{\Psi})\pi(\kappa_{n+k}|\kappa_{n+k-1},\boldsymbol{\Psi})\dots\pi(\kappa_n,\boldsymbol{\Psi}|\boldsymbol{y}_{1:n})\,
d\boldsymbol{\Psi}d\kappa_{n}\dots d\kappa_{n+k}$ and samples from it for forecasting.

%[[See p.534-535 of Pedroza paper; p.160-161 of Petris book]]

\section{IMPLICATIONS FOR ANNUITY PRICING}\label{sec:pricing}

In this section we aim to quantify the impact of longevity risk on the pricing
of annuities, using the mortality rates forecasted by the Lee-Carter
model in state-space form which is estimated by the Bayesian approach described in the
previous section.

\subsection{Estimation using Australian mortality data}

\begin{figure}[h]
\begin{center}
\includegraphics[width=7cm, height=5cm]{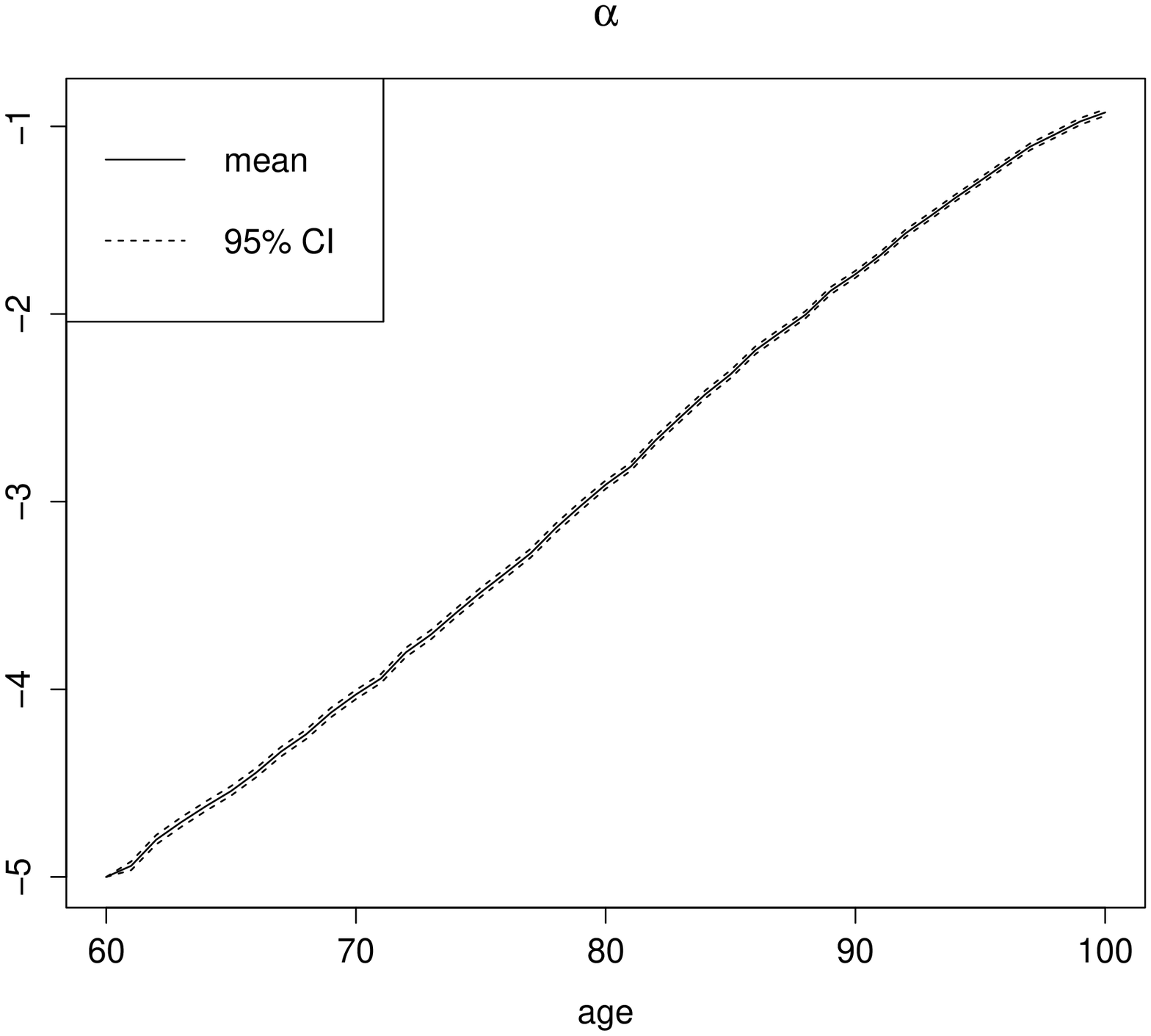}\includegraphics[width=7cm, height=5cm]{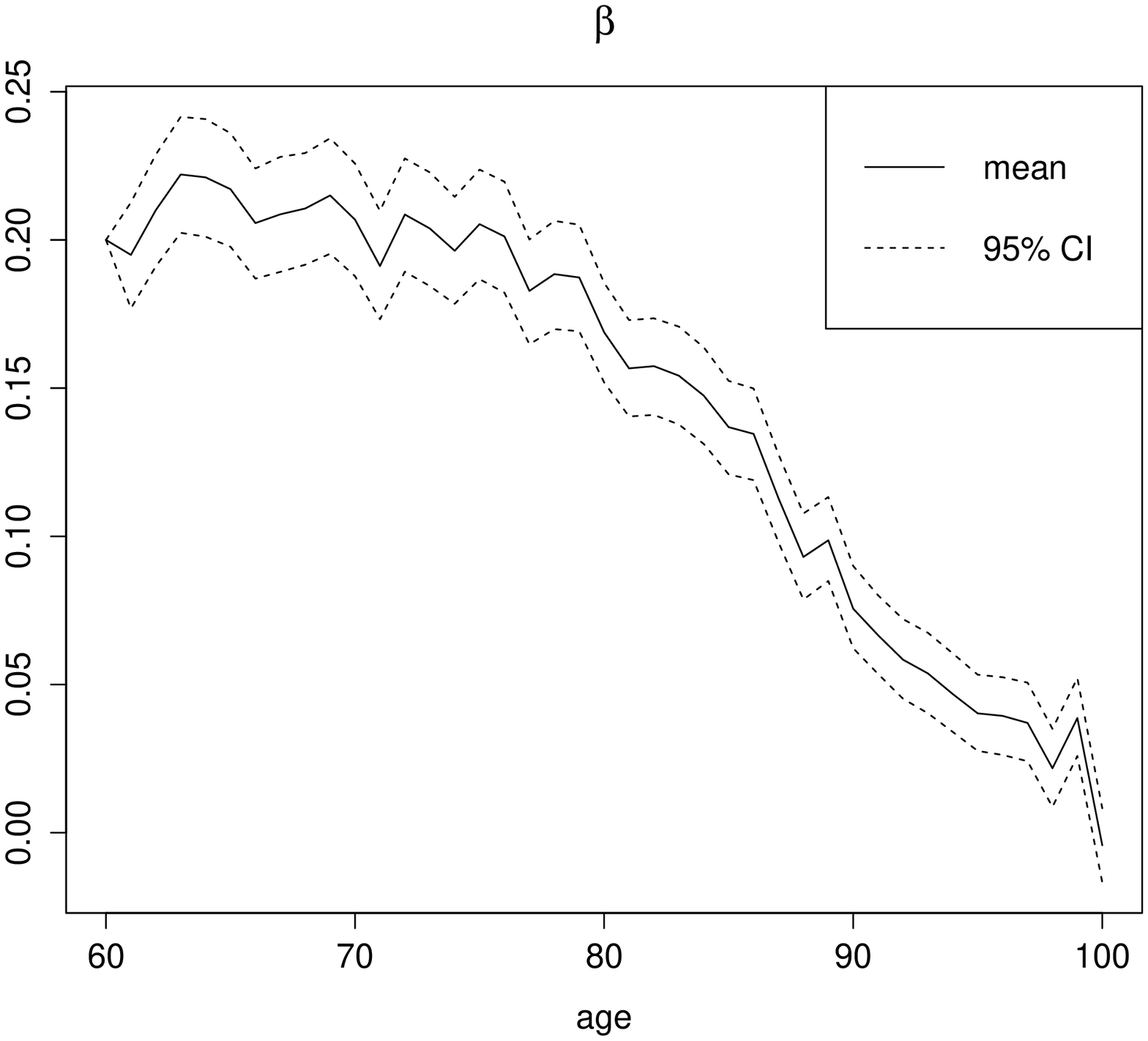}
\includegraphics[width=7cm, height=5cm]{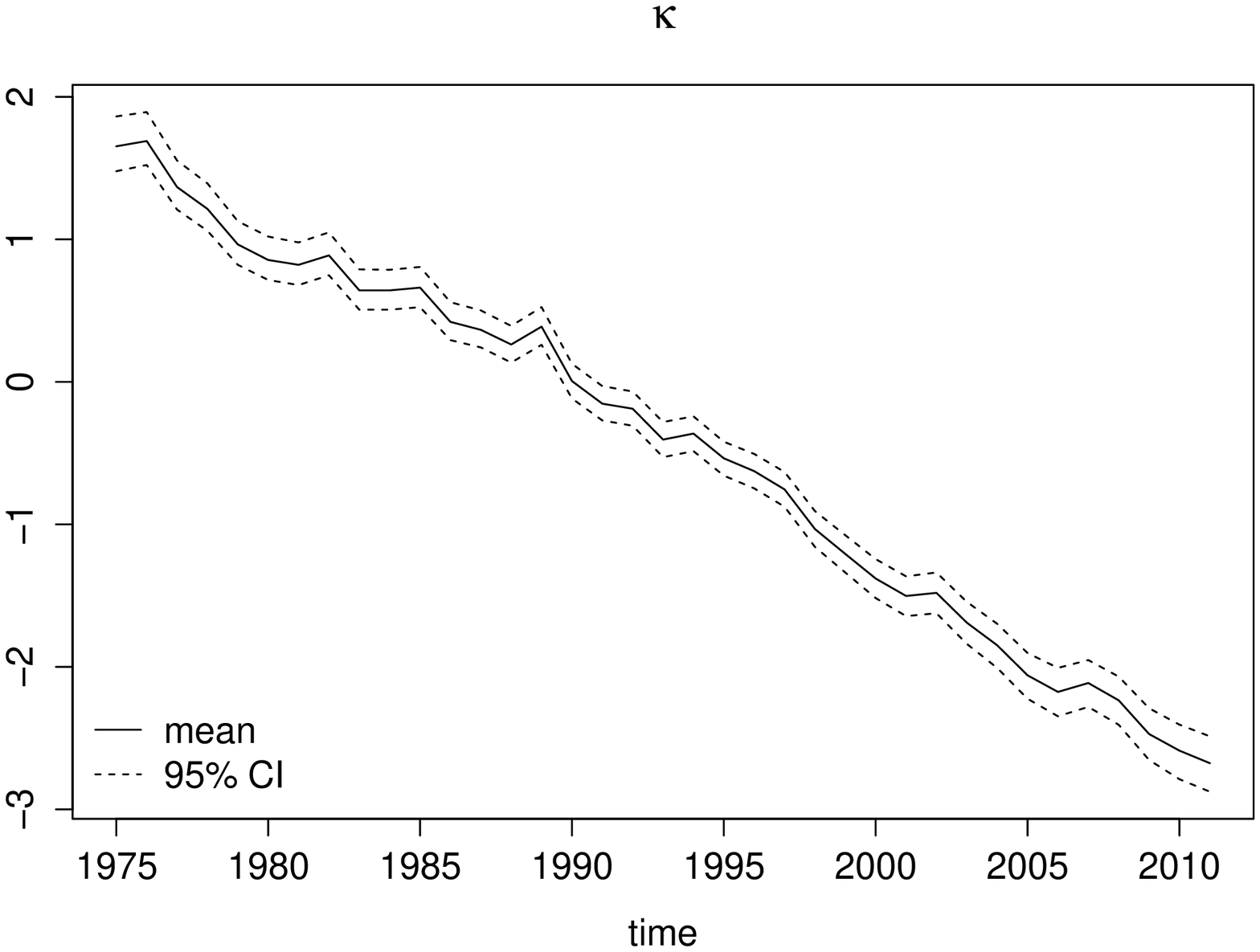}\includegraphics[width=7cm, height=5cm]{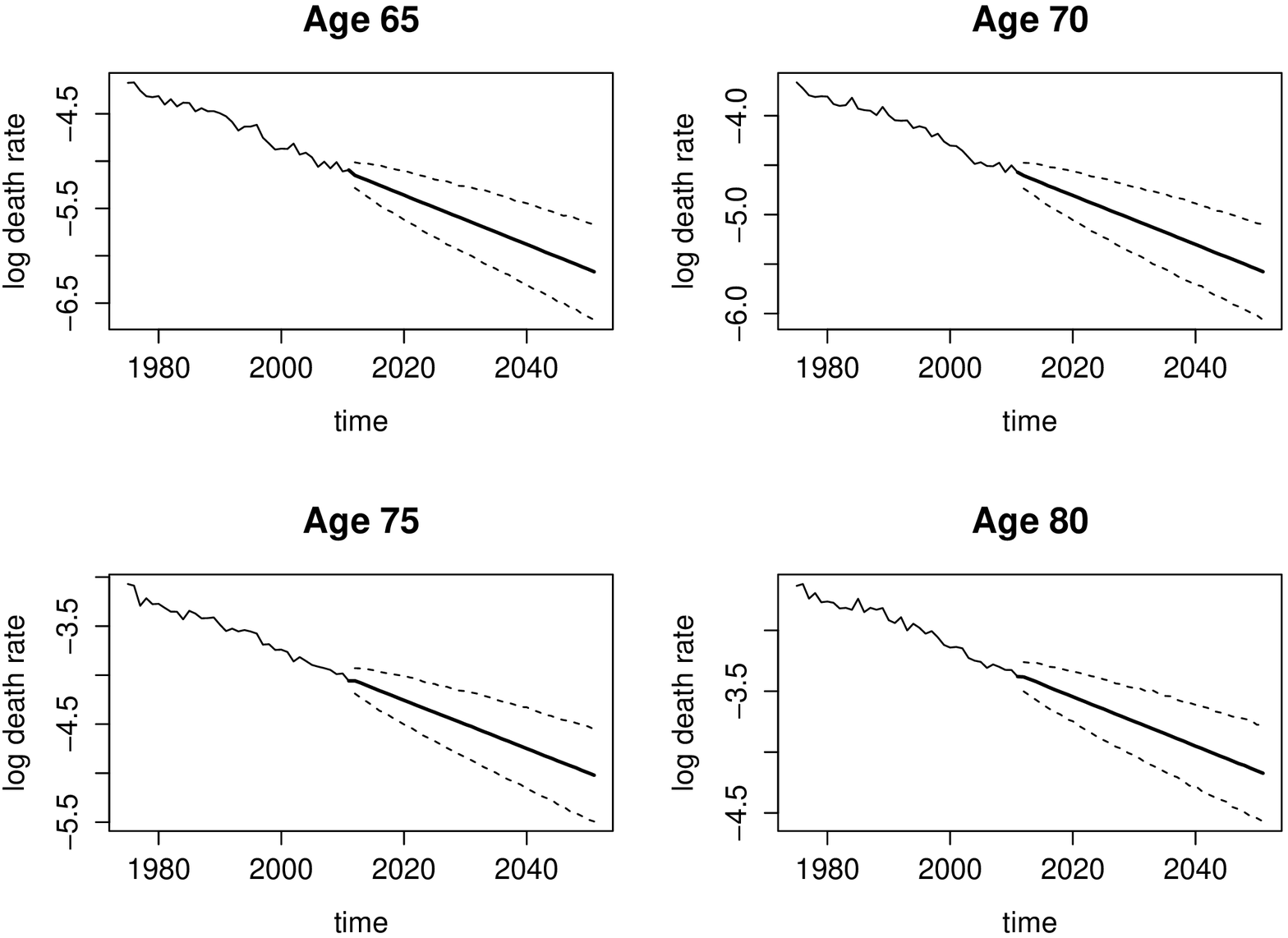}
\includegraphics[width=7cm, height=5cm]{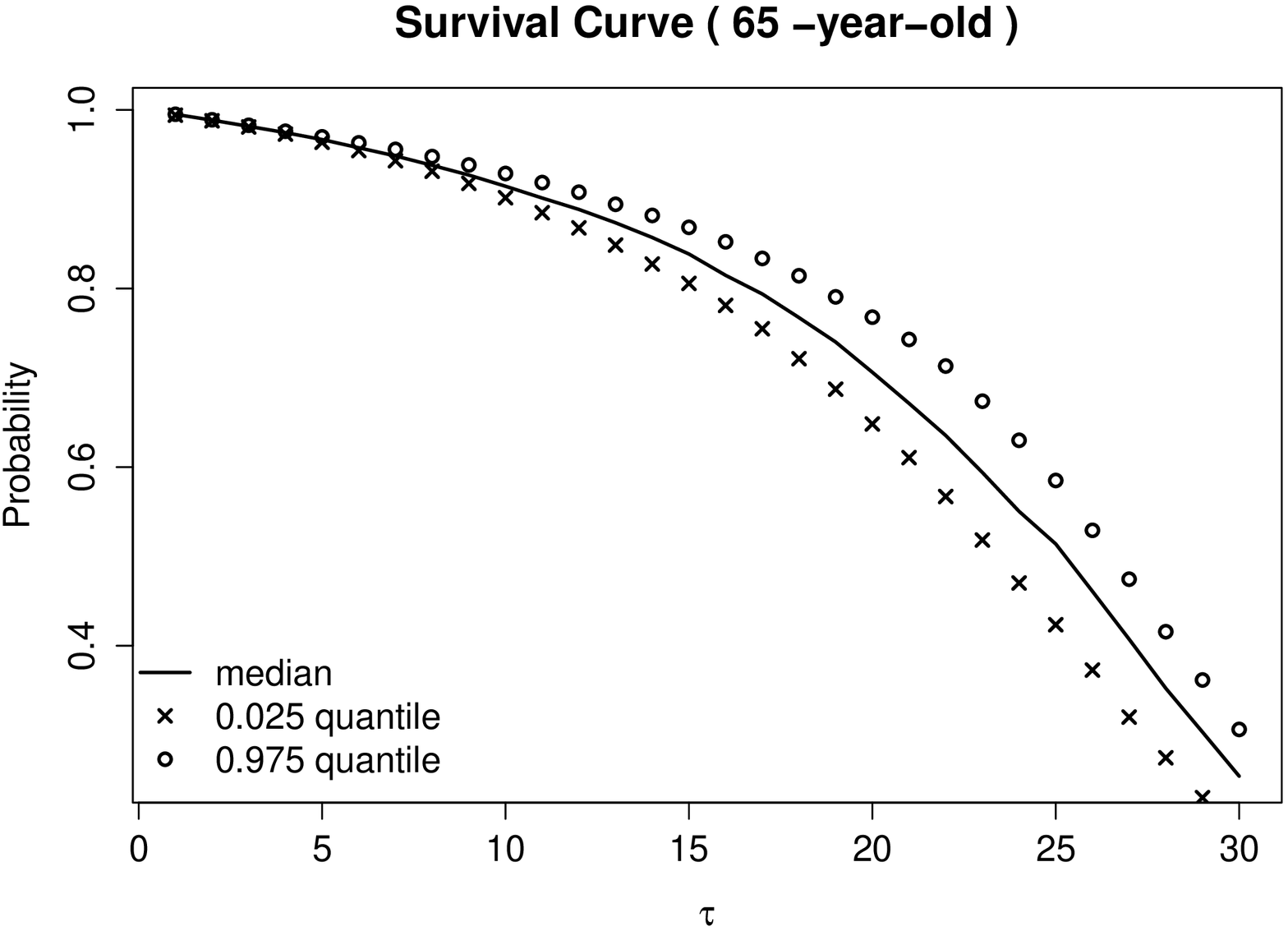}\includegraphics[width=7cm, height=5cm]{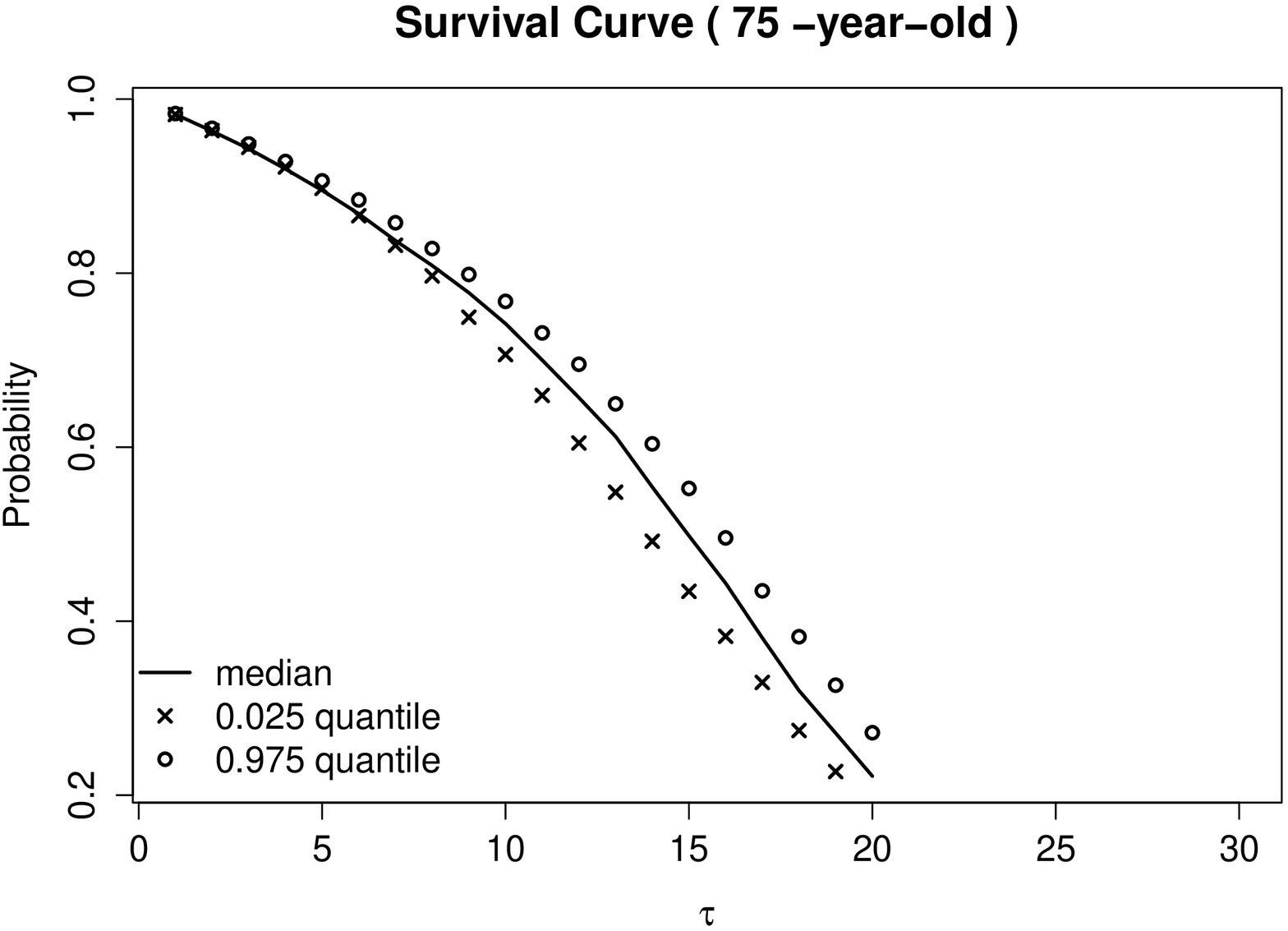}
\caption{(Upper four panels) Posterior mean and $95\%$ confidence interval (CI) for parameters $\boldsymbol{\alpha}$, $\boldsymbol{\beta}$; posterior mean and $95\%$ CI for the latent process $\kappa$ over year 1975-2011; mean and $95\%$ CI of the predictive distributions of log central death rates $(y_{65},y_{70},y_{75},y_{80})$ over 40 years forecast. (Lower two panels) Survival curves for different ages.}\label{fig:est}
\end{center}
\vspace{-1cm}
\end{figure}

The data set consists of Australian female mortality data obtained
from the Human Mortality Database (http://www.mortality.org). Since the application is for
annuity pricing, we focus on 1-year death rates for age 60-100 from
year 1975-2011. Figure~\ref{fig:est} shows the estimation results. Here we set $\alpha_{x_1}=-5$, $\beta_{x_1}=0.2$ and assume $m_0=0$, $C_0=100$ (these are the mean and variance of $\kappa_0$ used in Kalman filtering), $\tilde{\mu}_\alpha=\tilde{\mu}_\beta=\tilde{\mu}_\theta=0$, $\tilde{\sigma}^2_\alpha=\tilde{\sigma}^2_\beta=\tilde{\sigma}^2_\theta=100$, $\tilde{a}_\varepsilon=\tilde{a}_\omega=2.1$ and $\tilde{b}_\varepsilon=\tilde{b}_\omega=0.3$. Number of iterations in MCMC is $5000$ and the burn-in iterations is $1000$. We use very vague prior so that estimation is mainly determined by the data and the impact from prior is not material.

\subsection{Annuity pricing}

The $\tau$ year survival probability of a person aged $x$ currently (i.e. $t=0$ or year 2012) is determined by
\begin{equation}\label{eqn:survProb}
    {}_\tau p_x = \prod^\tau_{j=1}{}_1 p_{x+j-1}= \prod^\tau_{j=1}e^{-m_{x+j-1,j-1}}
\end{equation}
which is a random variable since $m_{x+j-1,j-1}$, for $j=1,\dots,\tau$, are random quantities forecasted by the Lee-Carter model (\cite{DenuitDh07}). Assuming a large enough annuity portfolio, the price of an annuity with maturity $T$ year, written for a $x$-year-old with benefit $\$1$ per year and conditional on the path $\boldsymbol{m}^x_{1:T}=(m_{x,0},m_{x+1,1},\dots,m_{x+T-1,T-1})$, is given by
\begin{equation}\label{eqn:annuity}
    a^T_{x}(\boldsymbol{m}^x_{1:T}) = \sum^{T}_{\tau=1} B(0,\tau)\,\text{E}(\mathsf{1}_{T_x>\tau}|\boldsymbol{m}^x_{1:\tau})= \sum^{T}_{\tau=1} B(0,\tau) {}_\tau p_x (\boldsymbol{m}^x_{1:\tau})
\end{equation}
where $B(0,\tau)$ is the $\tau$-year bond price, $\boldsymbol{m}^x_{1:\tau}$ is the first $\tau$ elements of $\boldsymbol{m}^x_{1:T}$, and ${}_\tau p_x (\boldsymbol{m}^x_{1:\tau})$ denotes the survival probability given $\boldsymbol{m}^x_{1:\tau}$ which is random. \cite{DenuitDh07} shows that some bounds of ${}_\tau p_x (\boldsymbol{m}^x_{1:\tau})$ can be computed analytically. \cite{Biffis05} evaluates annuity prices allowing for longevity risk using financial theory. From an annuity provider's perspective, what is important, however, is that the annuity price is a random quantity depending on the random paths of $\boldsymbol{m}^x_{1:T}$. Moreover, it is important to determine a survival curve ${}_\tau p_x$ (as a function of $\tau$) in \eqref{eqn:annuity} that best captures the mortality experience of the portfolio for risk management purposes. In this regard, we evaluate different quantiles of the annuity price $a^T_{x}(\boldsymbol{m}^x_{1:T})$ in Table~\ref{table:annuityprice} and extract the corresponding survival curves. Note that the forecasted death rate samples are used to produce sample paths $\boldsymbol{m}^{x,(\ell)}_{1:T}$ and hence samples of annuity prices $a^{T,(\ell)}_{x}(\boldsymbol{m}^x_{1:T})$ where $\ell=1,\dots,L$.  The bottom two panels of Fig.~\ref{fig:est} illustrates the survival curves corresponding to the median, 0.025 quantile and 0.975 quantile of the annuity price. Note that a less expensive annuity price indicates smaller survival probabilities.

\subsubsection{Impact of longevity risk}
The possibility that the realised survival curve would be different to the survival curve assumed for pricing leads to the so-called systematic mortality risk, a.k.a. longevity risk. In Table~\ref{table:annuityprice} we compare the median, 0.025 quantile and 0.975 quantile of the annuity prices for different ages and maturities. We also assume a constant interest rate $r=3\%$ and hence $B(0,\tau)=e^{-r\tau}$. Although the price difference might appear to be overall small, mis-pricing can be a significant risk when considering a large annuity portfolio. For an annuity portfolio consists of $N$ policies where the benefit per year is $B$, an under-pricing of $\gamma\%$ of the ``correct" annuity price will result in a shortfall of $NBa^T_x\gamma/100$ where $a^T_x$ is the ``wrong" annuity price being charged with benefit $\$1$ per year. For instance, $N=10,000$ policies written to 80-year-old policyholders with maturity $\tau=20$ years and $\$20,000$ benefit per year will result in a shortfall of $\$67$ million when the realised survival curve is the one that corresponds to the 0.975 quantile annuity price, while the survival curve corresponds to the median annuity price is assumed for pricing (here $\gamma=4.1$ in Table~\ref{table:annuityprice}). Moreover, as shown in Table~\ref{table:annuityprice}, mis-pricing is increasingly more pronounced for older ages as well as for annuity policies having a longer maturity.

\begin{table}[h]
\vspace{-0.5cm}
\center\small{\caption{Annuity price with different age and maturity ($T$) for female policyholder. Value in bracket ( ) is the percentage difference compared to median annuity price. We only consider contracts with maturity so that $\text{age}+\text{maturity} \leq 100$.}\label{table:annuityprice}}
\scalebox{0.9}{
\begin{tabular*}{1.0\textwidth}%
     {@{\extracolsep{\fill}}l|cccccc}
\hline
Maturity (years)  &  $T=5$ & $T=10$ & $T=15$ & $T=20$ & $T=25$ & $T=30$ \\
\hline
&\multicolumn{6}{c}{$\text{age}=65$}\\
\hline
Median   & 4.49          &  8.18          & 11.14          & 13.38          & 14.88          & 15.64 \\
0.025 Q  & 4.48 (-0.2\%) &  8.13 (-0.6\%) & 11.00 (-1.3\%) & 13.10 (-2.1\%) & 14.42 (-3.1\%) & 15.03 (-3.9\%) \\
0.975 Q  & 4.50 (+0.2\%) &  8.22 (+0.6\%) & 11.26 (+1.1\%) & 13.63 (+1.9\%) & 15.31 (+2.9\%) & 16.22 (+3.7\%) \\
\hline
&\multicolumn{6}{c}{$\text{age}=70$}\\
\hline
Median   & 4.42          &  7.94          & 10.57          & 12.30          & 13.15          & 13.41 \\
0.025 Q  & 4.41 (-0.4\%) &  7.86 (-1.0\%) & 10.37 (-1.9\%) & 11.92 (-3.1\%) & 12.63 (-4.0\%) & 12.82 (-4.4\%) \\
0.975 Q  & 4.44 (+0.4\%) &  8.01 (+0.9\%) & 10.76 (+1.8\%) & 12.66 (+2.9\%) & 13.67 (+4.0\%) & 14.00 (+4.4\%) \\
\hline
&\multicolumn{6}{c}{$\text{age}=75$}\\
\hline
Median   & 4.31          &  7.49          & 9.54          & 10.52          & 10.81          & N.A. \\
0.025 Q  & 4.29 (-0.7\%) &  7.38 (-1.6\%) & 9.27 (-2.8\%) & 10.12 (-3.8\%) & 10.35 (-4.3\%) & N.A. \\
0.975 Q  & 4.34 (+0.6\%) &  7.61 (+1.5\%) & 9.80 (+2.8\%) & 10.92 (+3.8\%) & 11.28 (+4.3\%) & N.A. \\
\hline
&\multicolumn{6}{c}{$\text{age}=80$}\\
\hline
Median   & 4.08          &  6.63          & 7.83          & 8.18          & N.A. & N.A. \\
0.025 Q  & 4.03 (-1.1\%) &  6.48 (-2.4\%) & 7.57 (-3.4\%) & 7.86 (-3.9\%) & N.A. & N.A. \\
0.975 Q  & 4.12 (+1.1\%) &  6.79 (+2.3\%) & 8.10 (+3.4\%) & 8.51 (+4.1\%) & N.A. & N.A. \\
\hline
\end{tabular*} }
\vspace{-0.5cm}
\end{table}

%\begin{table} \label{table:annuityprice}
%\caption{Annuity price for different age and maturity.}
%\begin{center}
%\begin{tabular}{cccc}
%\hline
%Break Proportion&Case 4&Case 5&Case 6\\
%\hline
%0&-1.612&-1.663&-1.665\\
%0.1&-1.645&-1.631&-1.646\\
%0.2&-1.600&-1.656&-1.652\\
%0.3&-1.590&-1.637&-1.654\\
%0.4&-1.600&-1.648&-1.668\\
%0.5&-1.618&-1.635&-1.644\\
%0.6&-1.553&-1.650&-1.686\\
%\hline
%\end{tabular}
%\end{center}
%\end{table}%

\section{CONCLUSIONS}
\label{sec:conclusion}
This article explores further the state-space representation of the Lee-Carter model in longevity modeling. We derive in details the posterior distributions of the static parameters and the latent process of the model under the Bayesian framework via Gibbs sampling. We suggest an identification constraint for the model that is particularly suitable for estimation under a MCMC approach. The predictive distributions of death rates are used to determine the range of annuity prices. Our results show that the assumption of survival curve has significant impact on annuity prices. Annuity written for older age policyholders is particularly vulnerable to mis-pricing caused by longevity risk. Extensions of the Lee-Carter model in state-space form and its estimation are currently under investigation.  % , even when the maturity of the annuity is relatively short.

\section*{Acknowledgement}
The research was supported by the CSIRO-Monash Superannuation Research Cluster, a collaboration among CSIRO, Monash University, Griffith University, the University of Western Australia, the University of Warwick, and stakeholders of the retirement system in the interest of better outcomes for all.

\bibliography{mcf}
\bibliographystyle{chicago} %use chicago style of referencing.
\end{document}